\documentclass[prl,twoside,onecolumn,superscriptaddress]{revtex4}
\usepackage[utf8]{inputenc}
\usepackage{graphicx}

\begin{document}

\title{Linear Polycatenanes from Kinetoplast Edge Loops}
\author{Josh Ragotskie}
\affiliation{Department of Physics and Astronomy, California State University, Long Beach}
\author{Nathaniel Morrison}
\affiliation{Department of Mathematics and Statistics, California State University, Long Beach}
\affiliation{Department of Physics, University of California, Berkeley}
\author{Christopher Stackhouse}
\affiliation{Department of Physics and Astronomy, California State University, Long Beach}

\author{Ryan C. Blair}
\affiliation{Department of Mathematics and Statistics, California State University, Long Beach}
\author{Alexander R. Klotz}
\affiliation{Department of Physics and Astronomy, California State University, Long Beach}
\affiliation{Corresponding author: alex.klotz@csulb.edu}

\begin{abstract}
    We use graph theory simulations and single molecule experiments to investigate percolation properties of kinetoplasts, the topologically linked mitochondrial DNA from trypanosome parasites. The edges of some kinetoplast networks contain a fiber of redundantly catenated DNA loops, but previous investigations of kinetoplast topology did not take this into account. Our graph simulations track the size of connected components in lattices as nodes are removed, analogous to the removal of minicircles from kinetoplasts. We find that when the edge loop is taken into account, the largest component after the network de-percolates is a remnant of the edge loop, before it undergoes a second percolation transition and breaks apart. This implies that stochastically removing minicircles from kinetoplast DNA would isolate large polycatenanes, which is observed in experiments that use photonicking to stochastically destroy kinetoplasts from \textit{Crithidia fasciculata}. Our results imply kinetoplasts may be used as a source of linear polycatenanes for future experiments.
\end{abstract}

\maketitle

\section{Introduction}

Catenanes are a form of mechanically interlocked molecule in which closed macrocycles are connected topologically like links in a chain. A polycatenane is a polymer chain in which the macrocycles serve as monomers and are polymerized by topological catenantion. Catenanes and polycatenanes are part of the growing field of mechanically interlocked molecules, which are of interest to the materials community because of the prospect of controlling physical properties, common chemistry and varied topology, and their potential use in nanomachines \cite{erbas2015artificial}. Polycatenanes have been produced synthetically \cite{wu2017poly}, and understanding the relationship between molecular topology and material properties is central to the development of catenated polymers. Single-molecule experiments are desirable for such a purpose: linear polymer physics has benefited from the use of viral genomic DNA molecules as model experimental systems \cite{marciel}, and it would be ideal if a similar system existed to study polycatenanes.

Kinetoplasts are complex DNA structures found in the mitochondria of trypanosome parasites. Often described as molecular chainmail, they consist of approximately 5,000 topologically linked circular molecules, known as minicircles, and several dozen larger molecules called maxicircles \cite{shapiro1995structure}. The kinetoplast is part of a complex gene editing apparatus: maxicircles contain the mitochondrial genes, which are transcribed into messenger RNA that must be edited by guide RNA molecules transcribed from the minicircles through the addition and removal of uracil \cite{sturm1990kinetoplast}. While the main reason for their study is to better understand and disrupt the parasite life cycle, in recent years they have seen use as a model system for two-dimensional polymer physics \cite{klotz2020equilibrium, soh2020deformation}. Typically it is their planar rather than their catenated nature that serves in this capacity, and it is unclear how their catenated molecular topology affects their mechanical properties, although more recent studies have begun to investigate this relation \cite{yadav2023tuning, he2023single}.

Most of what is known about kinetoplast topology comes from experiments on \textit{Crithidia fasciculata} (an insect parasite) kinetoplasts, in which minicircles were linearized with restriction enzymes and gel electrophoresis was used to determine the relative proportion of linear, circular, doubly-catenanted, and triply-catenated molecules \cite{chen1995topology}. It was concluded that each minicircle was connected to three others on average and that the regular lattice most consistent with the data was the honeycomb. High-resolution images of kinetoplasts appear to have disordered minicircles \cite{barker1980ultrastructure}, and a more recent analysis using atomic force microscopy supported the mean valence of 3, but not as a honeycomb lattice \cite{he2023single}. The maxicircles are catenated within the network and form their own catenating network \cite{shapiro1993kinetoplast}, but it is not known the extent to which they contribute to the overall structure or topology of the network. The relation between minicircle sequence and topology may not be random. For example, it is known that approximately 90\% of \textit{Crithidia fasciculata} minicircles have the same sequence, and a recent experiment examined the networks after removal of that majority (as well as the maxicircles)  using restriction enzymes \cite{yadav2023tuning}. The networks changed appearance but remained intact, suggesting that some combination the remaining 10\% of minicircles form a percolating network.

A lesser-known feature of kinetoplasts is that they are surrounded by a fiber of dense linked minicircles that serves a similar mechanical role as an elastic band in a shower cap. This edge fiber is of unknown biological function but becomes less prominent during kinetoplast replication. It is present in \textit{Crithidia fasciculata}, and has been observed across several trypansome genera including in \textit{Pythomonas serpens} (a plant parasite), \textit{Leishmania mexicana} (a mammal parasite), and \textit{Trypanosoma mega} (an amphibian parasite) \cite{barker1980ultrastructure} but does not appear to be prominent in the human-infecting parasites \textit{Trypanosoma brucei} and \textit{cruzi} \cite{guilbride1998replication}. The degree of redundancy, the number of parallel minicircles connecting hubs around the edge loop, is primarily estimated from electron or atomic force microscopy of kinetoplasts networks that have been spread on surfaces. The spreading procedures can degrade the networks, and as such these estimates must be treated as lower bounds. Barker \cite{barker1980ultrastructure} in 1980 stated that each hub contains 32 minicircles, with 16 minicircles connecting each hub to its neighbor on either side. In a more recent AFM study, He et al \cite{he2023single}. estimate based on the height of the hubs that they contain an average of 4.7 minircles. If two minicircles connect each hub to its two neighbors and one the interior, this suggests a mean redundancy of 2. It is not known whether the minicircles appearing on the edge have a specific sequence, although Kleisen et al. indicate that they are cut by the same endonucleases as the rest of the network \cite{kleisen}.

Experiments that remove minicircles from kinetoplasts will eventually remove enough such that the percolation threshold is passed. A network can be represented as a graph of nodes (in this case, minicircles) connected by edges (in this case Hopf linkages). If a small fraction of the edges between nodes are removed, a single connected component will still span the network. However, if a large fraction of edges are removed, the largest remaining connected component will be much smaller than the system size. The critical fraction of nodes such that a giant component exists is known as the percolation threshold, and depends on the underlying lattice topology \cite{van1997percolation}. In a kinetoplast experiment such as that of Chen et al. \cite{chen1995topology}, linearizing a minicircle is equivalent to removing all the edges connecting one node to its neighbors, and when the percolation threshold is passed, there should no longer be a cohesive DNA structure, simply an aggregate of formerly connected minicircles. If a kinetoplast has a purely honeycomb lattice topology (which has a percolation threshold of 70\%) the network should de-percolate when 30\% of minicircles are removed. Several of Chen et al.'s restriction assays were below this fraction, but the large connected components cannot pass through an agar gel and are not part of the data analysis. We note that in independent experiments from Ibrahim et al. \cite{ibrahim}, Li et al. \cite{tevin}, and Yadav et al. \cite{yadav2023tuning}, there is still DNA visible in the wells that likely corresponds to these remaining large clusters, with Li et al. presenting images of ``poly-interlocked'' structures. Beyond predicting the electrophoresis experiments for different lattices, theoretical work on kinetoplast topology has examined the probability that closely packed circles \cite{diao2015orientation} or ring polymers \cite{michieletto2015kinetoplast} will form a percolating network and the relationship between minicircle valence and percolation. No analysis to date (that we are aware of) has included the effect of the edge loop on percolation of a kinetoplast network. 

In this manuscript we examine percolation phenomena in networks with redundant edges, similar to the edge loop of kinetoplasts. Our purpose is to demonstrate that kinetoplasts may be useful source of linear polycatenanes for single molecular experiments. We support the findings from our simulations with fluorescence microscopy experiments in which photochemical cleavage is used to de-percolate \textit{Crithidia} kinetoplasts and show that the edge loops remain after the networks have largely been destroyed.

\begin{figure}
    \centering
    \includegraphics[width=0.9\textwidth]{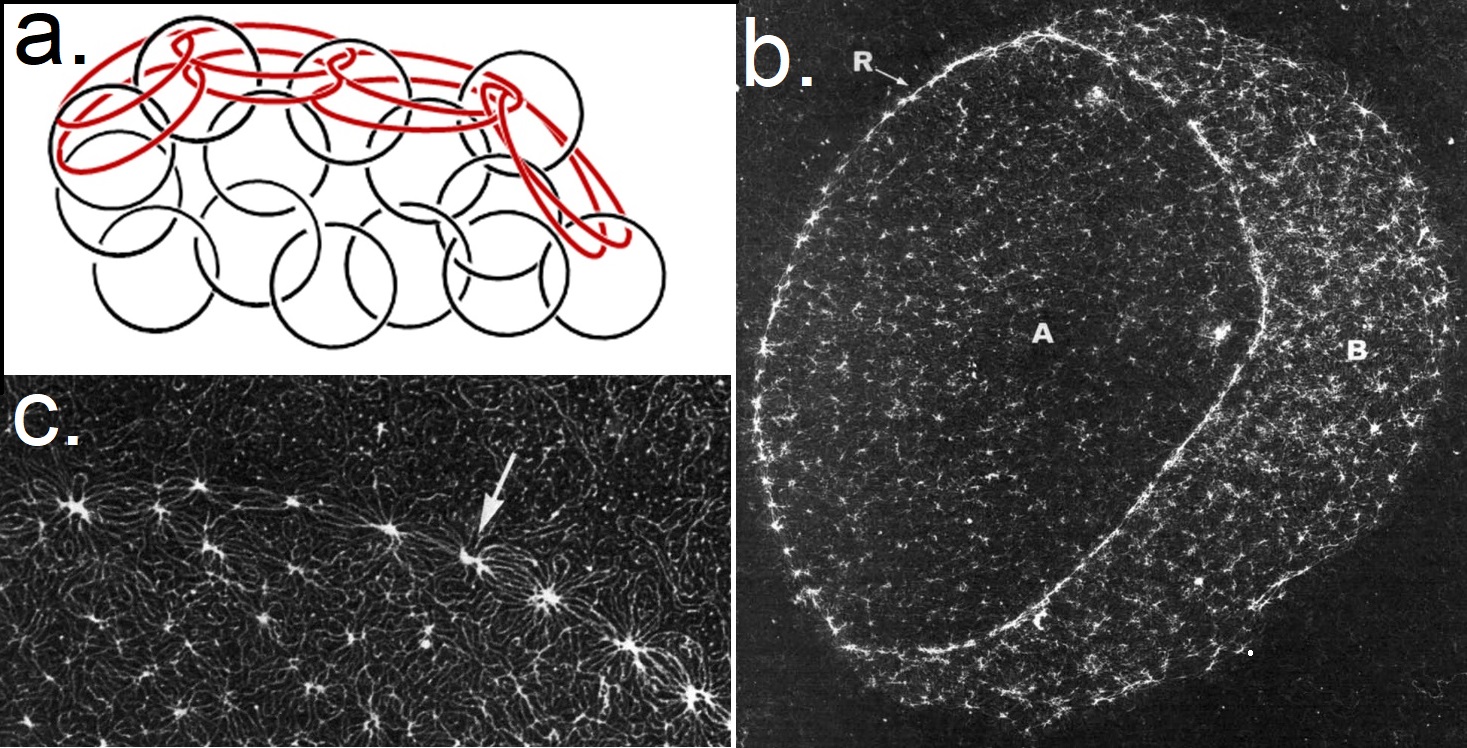}
    \caption{a. Schematic of the edge of a kinetoplast, in which a planar catenated DNA network is surrounded by a linearized band of redundantly linked minicircles. b. and c. show electron micrographs of the kinetoplasts from \textit{Crithidia fasciculata}, highlighting the edge loop and its features (used with permission from Barker, 1980 \cite{barker1980ultrastructure}). }
    \label{fig:fig1}
\end{figure}

\section{Simulations}

Following previous work on the topic \cite{chen1995topology,ibrahim,ferschweiler}, we represent kinetoplast networks as graphs in which each node represents a minicircle, and an edge between two nodes represents a topological connection between two minicircles. The removal of a node from a network is equivalent to breaking a minicircle, either by restriction enzyme digestion or photonicking. If a subgraph remains but all nodes surrounding it are removed, the subgraph is no longer connected to the rest of the network. The computational portion of this work involves measuring the size of connected subgraphs in networks from which nodes have been removed.

%When the fraction of nodes in a network exceeds a critical value, there exists a single giant connected component that spans the entire lengthscale of the network. This critical value is known as the percolation threshold, below which a giant connected component does not exist. For example, the percolation threshold of the square lattice is approximately 0.59, meaning that if 40 \% of nodes are removed at random from a square lattice, there will still be a giant component, but if 42 \% of nodes are removed it will break into many small fragments.

As is standard in graph theory \cite{graph}, we represent lattices as their corresponding adjacency matrices. The adjacency matrices for undirected, unweighted lattices are symmetric binary matrices where each node is represented as both a row and a column, and a non-zero entry in row $x$ and column $y$ indicates node $x$ shares an edge with node $y$. To mimic the effect of the edge loop, we assign nodes on the edges an integer value $R\geq 1$  representing the number of parallel minicircles. Similarly, the non-zero entries in the rows and columns of the adjacency matrix corresponding to nodes on the edge are all $R$. Removing a node is equivalent either to removing the corresponding row and column from the adjacency matrix or to reducing the non-zero labels in row and column of an edge node from $R$ to $R-1$. For example, if $R=5$, then the same node would have to be reduced five times in order to ``cut'' it completely. To investigate broad features of redundantly-edged lattices, we initialize networks of various topology and simulate the removal of minicircles by zeroing nodes in the network. We determine which occupied nodes are connected in the same subgraph, to calculate a list of the sizes of all the connected components in the system.

\begin{figure}[h]
\centering
\includegraphics[scale=.37]{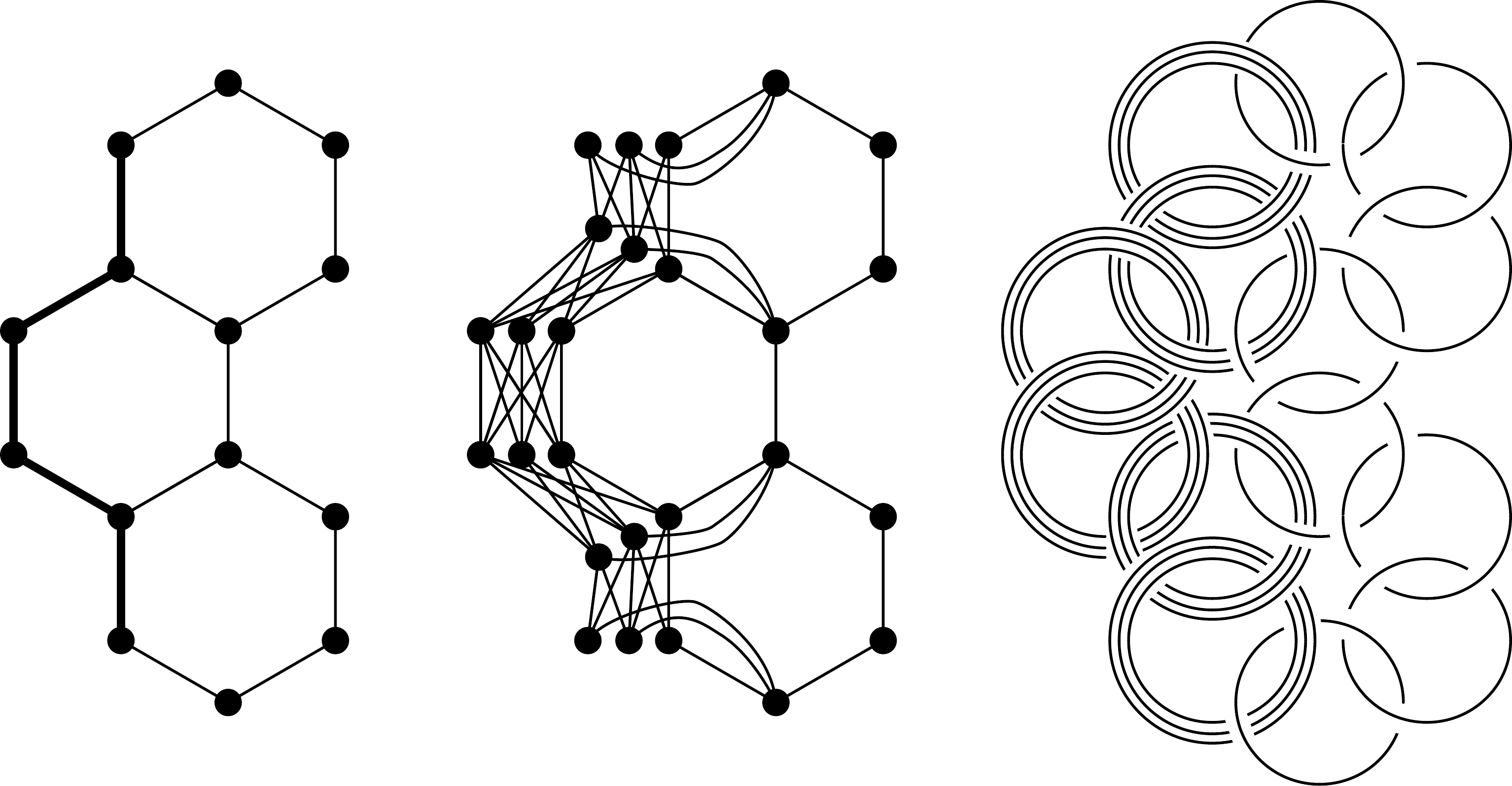}
\caption{Right: A portion of the kinetoplast in the hexagonal model near the boundary where the boundary is modeled by 3-fold catenated parallel loops. Center: The graph encoding the kinetoplast network for this portion of the kinetoplast. Left: The simplified graph encoding the kinetoplast network where the boundary of the portion of the planar honeycomb lattice is indicated with thickened edges.}
\label{fig:BoundaryExample}
\end{figure}

%To mimic the effect of the edge loop, we represent lattices as binary matrices but give nodes on the edges an integer value $R\geq 1$  representing the number of parallel minicircles. Removing a node is equivalent either to reducing a matrix element in the interior of the network from 1 to 0, or to reducing an edge node's value by 1. For example, if R=5 then the same node would have to be reduced five times in order to ``cut'' it. To investigate broad features of redundantly-edged lattices, we initialize networks of various topology and simulate the removal of minicircles by zeroing nodes in the network. We determine which occupied nodes are connected in the same subgraph, to calculate a list of the sizes of all the connected components in the system.

We investigate square, honeycomb, and ``random'' lattices. Random lattices are generated from random points on a sparse square lattice and choosing a common radius $r$ such that the probably of two nodes sharing an edge is proportional to the overlap area between two circles of radius $r$ centered on each node. The radius is chosen such that the mean valence of each node is approximately 3, consistent with what is expected of kinetoplast DNA.

Degradation of the kinetoplast is modeled by an iterated process of randomly removing a vertex, and all edges incident to that vertex, from a spatial graph. We call such a removal a dissolution. After each dissolution step, the size of every connected component in the graph, down to individual isolated vertices, is recorded. This process proceeds until every connected component of the resulting graph has fewer vertices than a threshold value. For the computations discussed in this paper, we chose this threshold value to be 100, from an initial ~5000.

\subsection{Simulation Results}

Many properties of percolating network manifest as size-independence or apply in the limit of very large networks when edge effects are neglibible. However, because we are studying biological networks with approximately 5000 components, and because we are explicitly studying an edge effect (that of the catenated loop), we are not concerned with size-independence or the asymptotically large limit. To identify percolation thresholds in our simulations, we measure the second-largest cluster of the networks and observe its local maximum \cite{secondlargest}. While this metric has weak dependence on the system size, it is suitable for our investigations.

\begin{figure}
    \centering
    \includegraphics[width=0.9\textwidth]{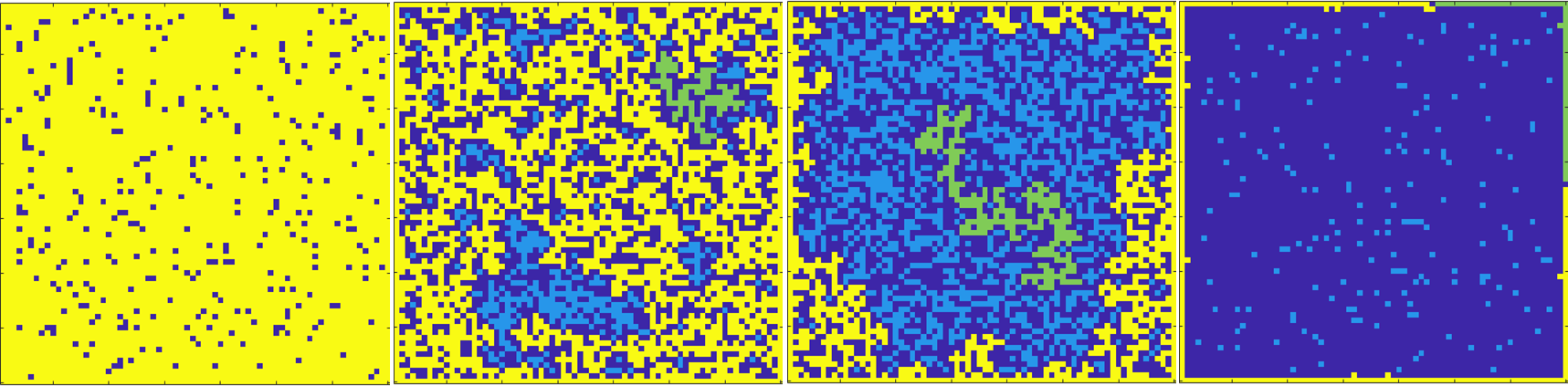}
    \caption{The dissolution of a 70x70 square lattice network with ten-fold edge redundancy. The largest connected component is labelled in yellow, the second largest in green, other components as light blue, and unoccupied sites in dark blue. The first image shows a giant component with a few removed nodes. The second is at the percolation threshold of the network interior, in which the largest component still spans the entire system. In the third image, the interior is no longer percolating but the edge of the system still contains its largest component. In the fourth image, the edge has broken into two components.}
    \label{fig:squaredraw}
\end{figure}

The salient features of our model can be seen in the two supplemental videos, which show the dissolution of two 70x70 networks on a square lattice, one without edge redundancy and one with R=10, with snapshots shown in Fig. \ref{fig:squaredraw}. The largest component is displayed in yellow, the second largest (when unique) in green, other components as light blue, and unoccupied sites in dark blue. As each video plays, the charts below show the largest component size in blue and the second-largest in red, scaled up for visibility. In the edgeless network, the largest component ceases to span the network near the expected percolation threshold and then rapidly decreases in size. In the edge-redundant network, after the interior percolation transition, the largest component remains localized to the perimeter of the lattice, until it begins to break into many smaller fragments at the second percolation transition, at which point both the largest and second-largest component lie on the edge. Experimentally, this would manifest as several larger linear polycatenanes amidst a sea of smaller fragments.

\begin{figure}
    \centering
    \includegraphics[width=1\textwidth]{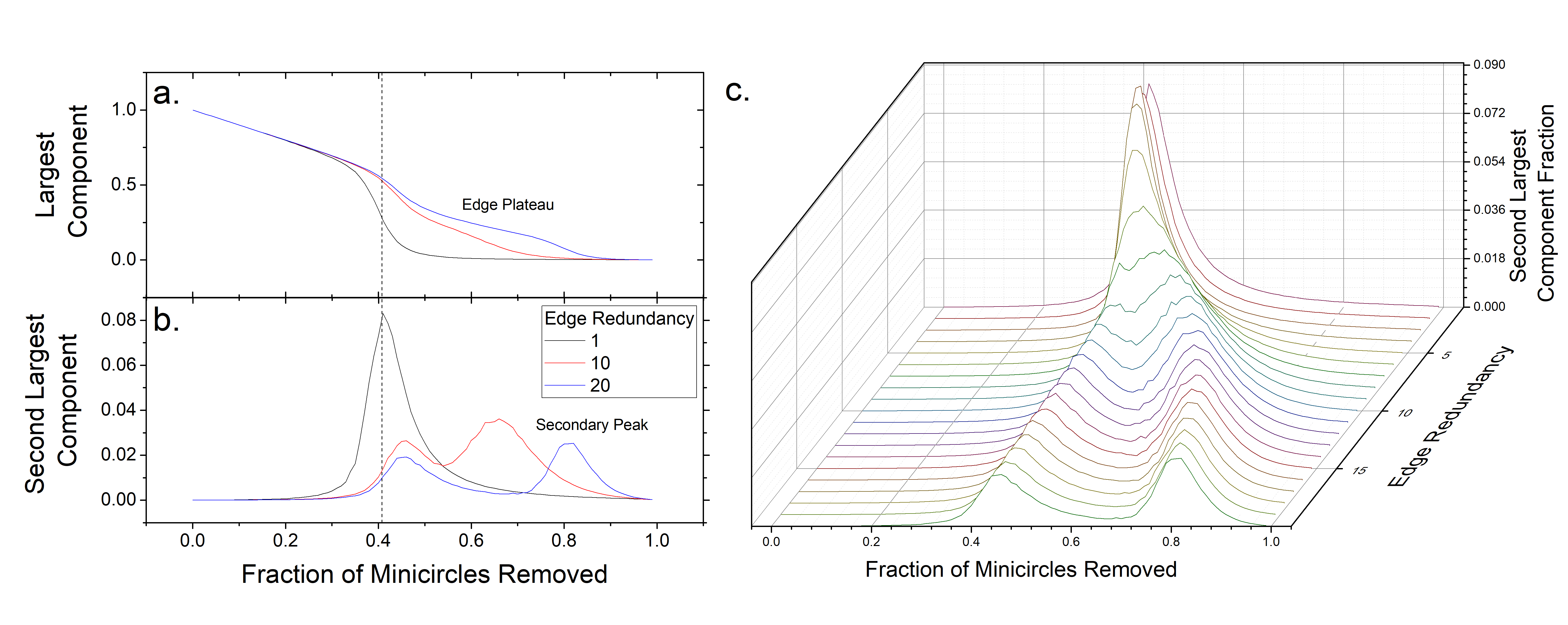}
    \caption{a. The size of the largest connected component in 70x70 square networks with edge redundancy of 1, 10, and 20 as a function of the fraction of dissolved minicircles. At a dissolution fraction that leaves an edgeless lattice with a small largest component, the edged lattices have a larger largest component around their edge. b. The size of the second largest connected component in 70x70 square networks with edge redundancy of 1, 10, and 20. A local maximum indicates a percolation transition. With a redundant edge, a secondary transition arises when the edge breaks. The dashed line shows the percolation threshold of a square lattice. c. Waterfall plots of the second largest components for lattices with edge redundancy between 1 and 20.}
    \label{fig:deperc}
\end{figure}

The basic features of edged network percolation can be seen in Fig. \ref{fig:deperc}, where data is presented from a square lattice simulation. For a network with no redundant edge, below the percolation threshold (corresponding to the removal of 0.4073 of nodes) the largest component is small compared to the entire lattice. With redundant edges, there remains a significantly sized component around the circumference of the network, leaving a largest component taking up a non-negligible fraction of the network. The percolation threshold may be identified by a local maximum in the second-largest component. This is observed at the expected value for edgeless networks,  but as the edge redundancy increases a secondary peak emerges corresonding to a second percolation transition, in which the edge breaks up into two or more components. The waterfall plot shows this process as edge redundancy increases, increasing the amplitude of the secondary percolation peak and pushing it towards a higher fraction of removed minicircles. This implies, for example that when 70\% of minicircles are removed, an edgeless network will have no large components, whereas a network with a sufficiently redundant edge will still have a largely pristine edge fiber with an empty interior.

\begin{figure}
    \centering
    \includegraphics[width=0.9\textwidth]{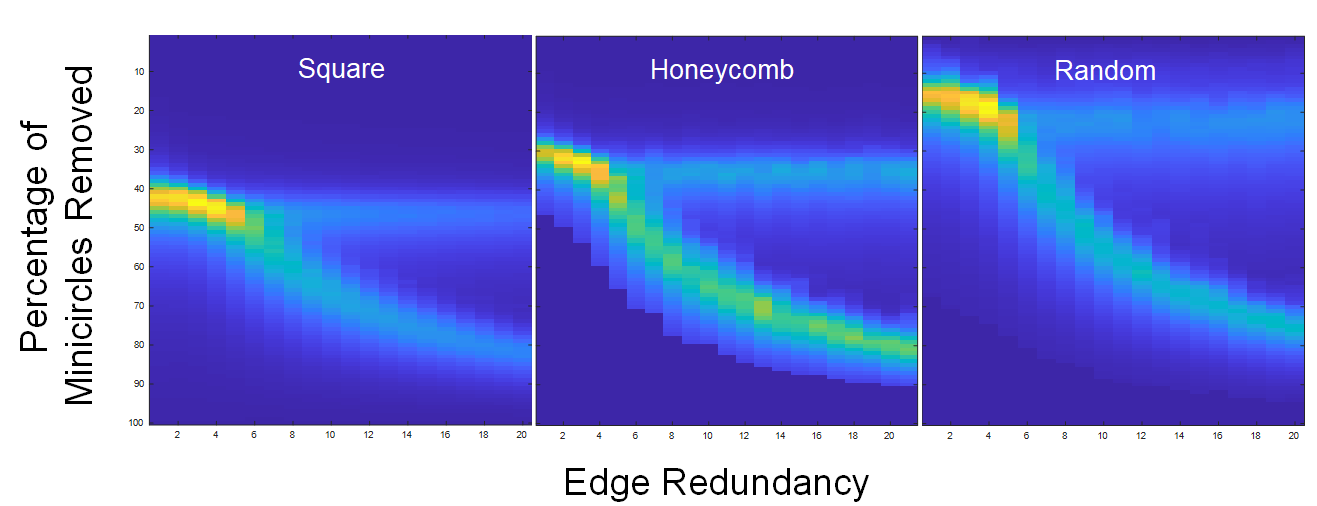}
    \caption{Heatmaps showing the second largest cluster size as a function of the percentage of minicircles removed (from top to bottom) and edge redundancy up to 20 (left to right) for square, honeycomb, and random networks. In all cases, edge redundancy causes the peak to split into two, the upper conrresponding to interior de-percolation and the the lower corresponding to edge loop fragmentation.}
    \label{fig:latticecomp}
\end{figure}

We can compare the effect of lattice geometry by examining heatmaps showing the effect of increasing edge redundancy on the second largest cluster size at different removal fractions. Heatmaps in Figure \ref{fig:latticecomp} show the second largest component size as a function of a the fraction of removed minicircles and the edge redundancy, for the square, honeycomb, and random networks. The known percolation thresholds for the square and honeycomb lattices (1-0.59=0.41 and 1-0.70=0.3 respectively) are recovered for the edgeless lattices, and the random lattice has a higher percolation threshold. The higher percolation threshold for the random lattice is likely due to the topological similarities between the random lattice and lattices with defects, which have previously been shown to have a higher percolation threshold \cite{spencerziff}. As the redundancy increases, the peak splits into two with the top peak, corresponding to interior de-percolation, remaining at its edgeless location and the second peak, corresponding to edge fragmentation, being found at a removal fraction that increases with redundancy.

\begin{figure}
    \centering
    \includegraphics[width=0.9\textwidth]{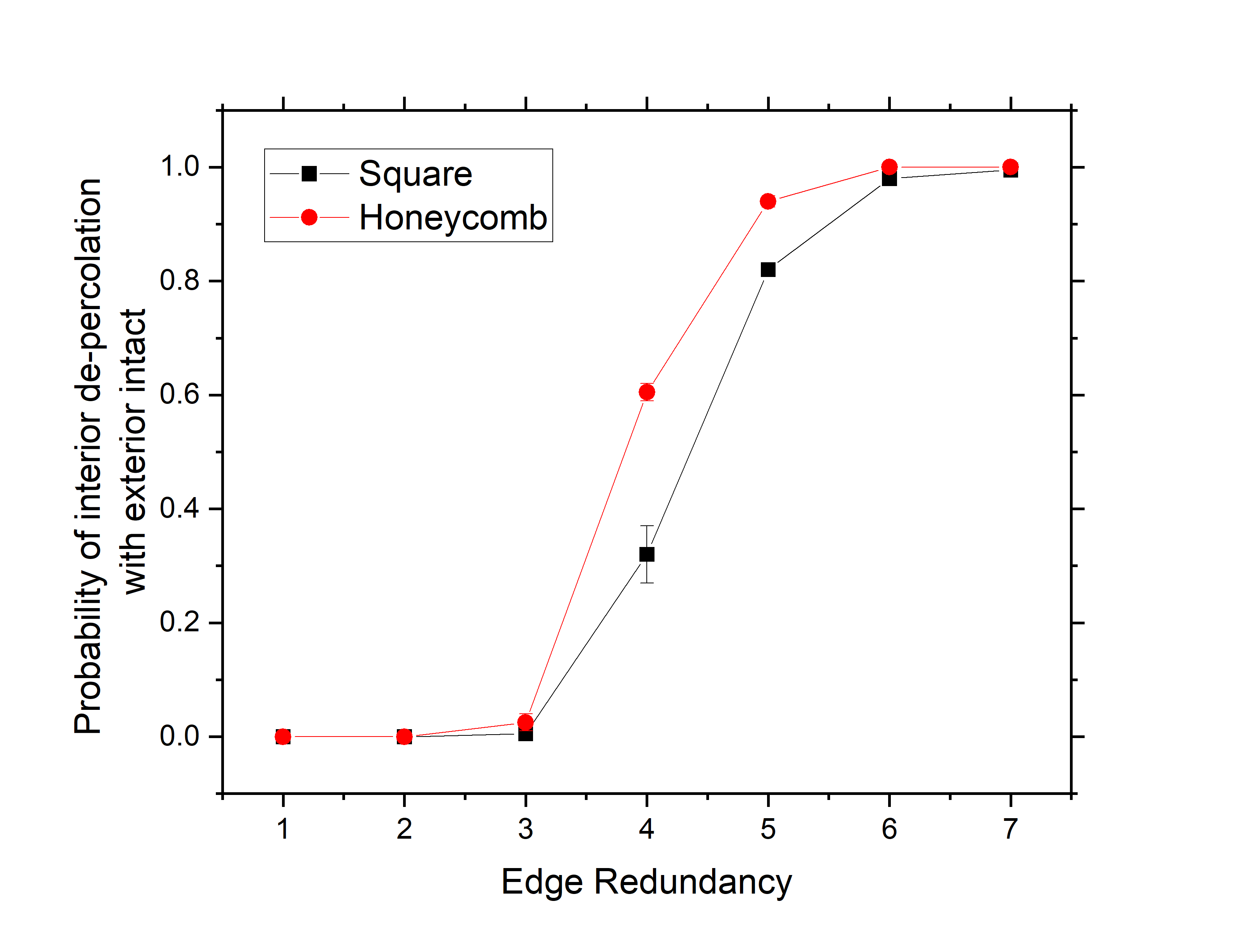}
    \caption{Probability that the interior of a network de-percolates before the exterior breaks, as defined in the text, as a function of the edge redundancy for a 70x70 square lattice and a 5000-node honeycomb lattice. When not visible, standard error bars are smaller than points.}
    \label{fig:deperc2}
\end{figure}

Finally, to support our assertion that edge redundancy leads to the edge loop being the last cohesive component of the network, we can vary the edge redundancy, perform dissolution simulations, and measure whether the interior de-percolates before the exterior breaks. Interior de-percolation is defined as follows: a graph consisting of every non-peripheral node is analyzed for connected components, and the coordinates of the largest component are analyzed to determine whether it touches the edge on all four sides. If it does not, the interior is considered de-percolated. The exterior is determined to have broken if the value of any matrix element on the edge reaches zero during the dissolution process. In Figure \ref{fig:deperc2} we plot the probability that the interior de-percolates before the exterior breaks, as a function of the edge redundnacy for a square lattice with 4900 nodes and a hexagonal lattice with 5000 nodes. For $R\leq3$ it is very unlikely for the edge to survive until the interior de-percolates, whereas for $R\geq6$ it is almost guaranteed that the edge will be intact when the interior depercolates. 

\section{Experiments}

\begin{figure}
    \centering
    \includegraphics[width=0.9\textwidth]{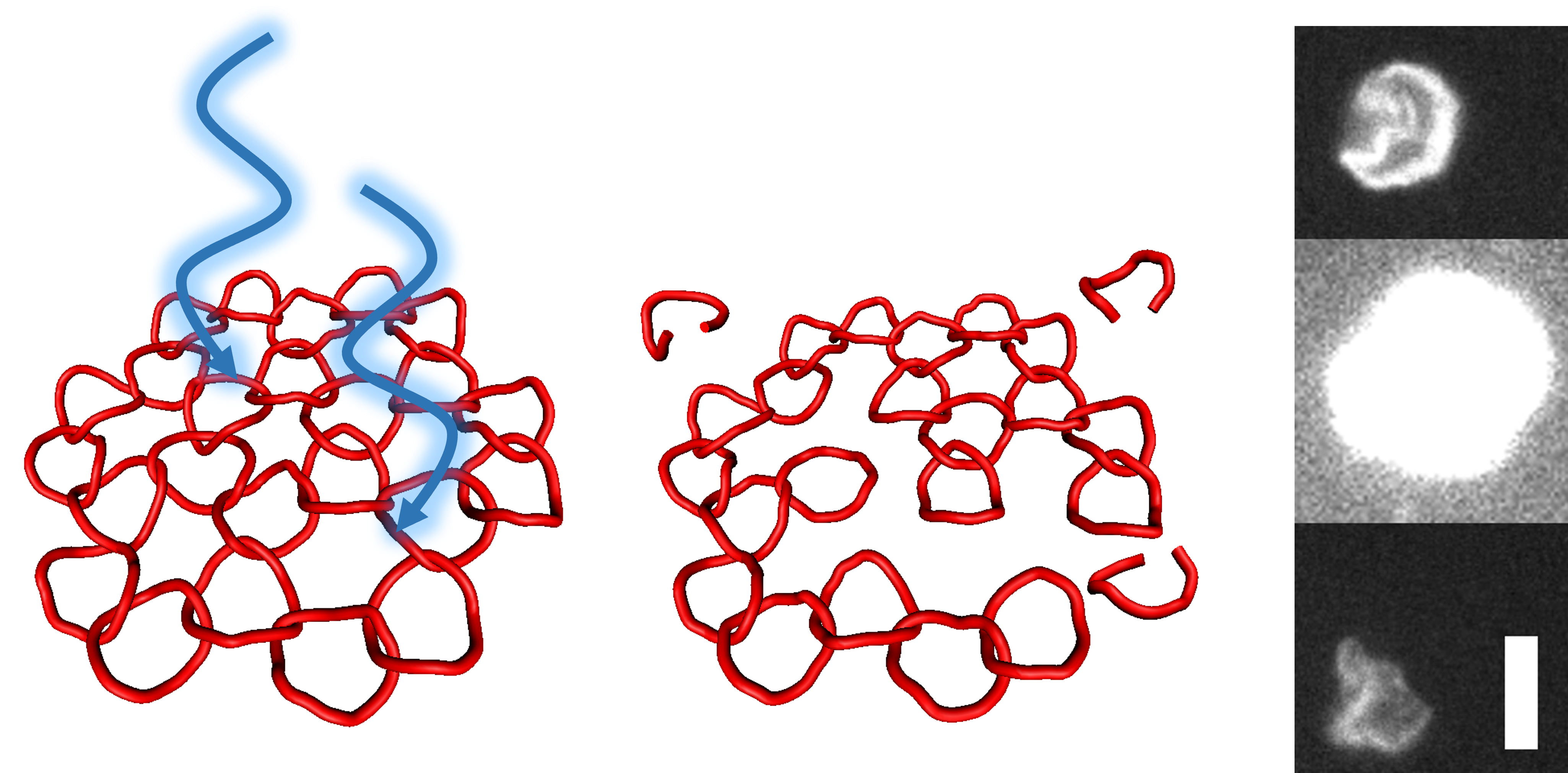}
    \caption{Schematic of our experiments in which intense light is used to linearize minicircles and remove them from the network. Right side shows an image of a kinetoplast before, during, and after the light intensity is increased, with 5 micron scale bar. }
    \label{fig:method}
\end{figure}

Our experiments are meant to demonstrate that fibers from the edge loop are the last cohesive remnants of destroyed kinetoplasts. We visualize kinetoplasts from \textit{Crithidia fasciculata} (Topogen, Inc.) using fluorescence microscopy by staining them with YOYO-1 dye at a ratio of 4 base-pairs per YOYO in 0.5x TBE buffer with 4\% beta-mercaptoethanol according to procedures described previously \cite{klotz2020equilibrium}. The kinetoplast-containing buffer are sealed between two glass cover slips with 100-micron thick double-sided tape. The molecules are illuminated with an LED filtered to 490 nm (blue) and visualized with a Leica DMI8 inverted fluorescence microscope. The combination of YOYO-1, DNA, light, and dissolved oxygen can induce a reaction that breaks DNA \cite{aakerman1996single}. Typically this is undesirable and is prevented by the additon of beta-mercaptoethanol or glucose oxidase to the buffer, by degassing buffer solutions to remove dissolved oxygen, and by limiting the intensity of the illumination \cite{persson2010dna}. To promote photodisintegration of the DNA we forewent the degassing phase and increased the illumination intensity far beyond the recommended amount (Fig. \ref{fig:method}), which causes minicircles to rapidly break. The mechanism of YOYO-1 photocleavage of DNA, which happens when two single-stranded breakages occur opposite one-another, is discussed by Akerman and Tuite \cite{aakerman1996single} and examined on a single-molecule level by Alazadehheidari et al. \cite{alizadehheidari2015nanoconfined}.

\begin{figure}
    \centering
    \includegraphics[width=0.9\textwidth]{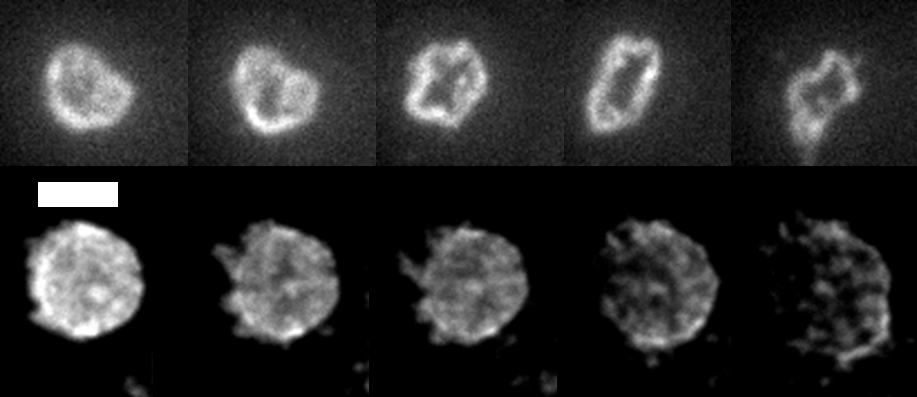}
    \caption{Two montages of kinetoplasts undergoing photodisintegration as their interior de-percolates. The top montage shows a free kinetoplast which loses its interior connectivity, resulting in a thick circular polycatanane, which continues to degrade. The bottom montage shows a kinetoplast adsorbed to a surface, in which visible holes appear in the interior. Eventually the edge loop breaks and the kinetoplast opens up.}
    \label{fig:breakage}
\end{figure}

\begin{figure}
    \centering
    \includegraphics[width=0.9\textwidth]{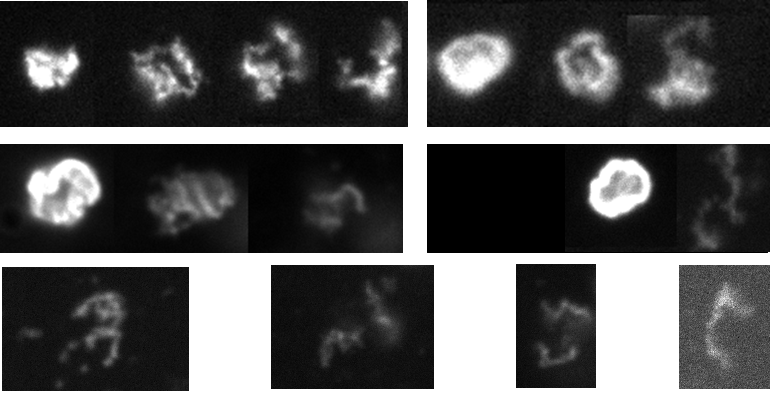}
    \caption{Four montages of kinetoplasts undergoing photodisintegration, leaving behind a fibrous polycatenane. The bottom row shows the end stage of four more degradations.}
    \label{fig:breakage2}
\end{figure}

Figure \ref{fig:breakage} shows two time lapses of a molecule undergoing photodisintegration. The top montage shows a molecule in free solution that forms a visible hole in its interior, leaving a contiguous boundary. Eventually the boundary becomes more like a thick ring polymer, undergoing shape fluctuations that are not observed in whole kinetoplasts \cite{klotz2020equilibrium}. It has been shown that kinetoplasts with a significant fraction of removed minicircles (while still forming a percolating network) exhibit larger-amplitude fluctuations \cite{yadav2023tuning}, which we confirm. The bottom montage shows a kinetoplast absorbed to a glass over slip. Dark patches appear in the interior and become more prominent until the edge loop breaks and the molecule opens up, leaving a large edge loop fragment and a distribution of smaller interior fragments. Because kinetoplasts can be considerably larger than the focal plane of a high numerical aperture lens, it can be difficult observing the whole process with such clarity. Several more time of photodisintegration lapses can be seen in Fig. \ref{fig:breakage2}, as well as some other examples of edge loops post-disintegration.

\begin{figure}
    \centering
    \includegraphics[width=0.9\textwidth]{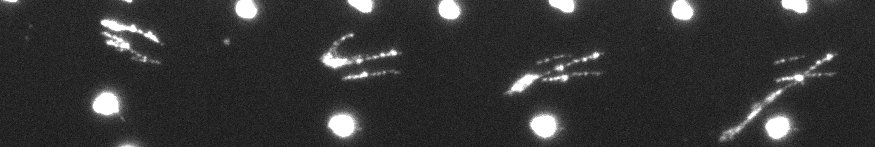}
    \caption{Montage from an experiment in which a kinetoplast is trapped in a gel network under an applied field while being photodisintegrated. The edge loop at this point is the only remaining cohesive structure and becomes significantly stretched. }
    \label{fig:gel}
\end{figure}

We attempted to observe the dissociated edge loops undergoing agar gel electrophoresis with optical microscopy. We performed experiments in which 400 $\mu$m tall microfluidic channels (Ibidi GmbH) were filled with an agarose solution which was allowed to gelify. Kinetoplasts were placed in the microchannel reservoir and driven with an electric field towards the boundary of the gel where their motion was arrested. After photodistintegration, the remaining edge loops did not penetrate the gel. However, because the gel typically contracted over the course of the experiment, there were continuous gel domains separated by regions of fluid through which kinetoplasts could pass. This often lead to situations in which kinetoplasts were trapped between regions of gel with the electric field straining them in unexpected directions. As kinetoplasts began to photodisintegrate, their released fragments would translate away from them, and eventually their edge fibers would be liberated from the interior and stretch dramatically as in Figure \ref{fig:gel}. The circumference of a kinetoplast is approximately 16 microns, comparable to the maximum observed length of the stretched fiber.

\section{Discussion and Conclusions}

We have investigated the effect of the kinetoplast edge loop on the percolation properties of the network under random minicircle removal. We sought to investigate the effect of the edge loop on the kinetoplast's percolation behavior, and found that the edge loop or fragments from it are likely to be the last large remaining component when a kinetoplast is destroyed by minicircle removal. To verify this, we destroyed kinetoplasts using YOYO-1 photocleavage and observed similar features as those predicted in simulations. This suggests that stochastically degraded kinetoplasts can serve as a source of polycatenanes for future polymer experiments. It was not our goal in this work to characterize the equilibrium or elastic properties of these isolated edge loops, which is a topic of future study.

More precise measurements of the percolation threshold of \textit{Crithidia fasciculata} may provide insight into the network's topology. In gel electrophoresis experiments, the non-penetrating fraction of DNA may be collected and visualized to determine if they primarily consist of holey membranes (as in Yadav et al. \cite{yadav2023tuning}) or edge loops (as in this work), and the disappearance of the non-penetrating segment may serve as evidence that the second percolation threshold, that which breaks the edge loop, has been surpassed. As seen in Fig. \ref{fig:deperc}, the first percolation threshold depends on the interior lattice topology while the second depends on the degree of edge redundancy. Thus, an experiment that measures the enzyme activity at which each threshold occurs could provide a new estimate of both the lattice topology and the edge redundancy. Finally, our simulations that suggest an edge redundancy of at least 4 is required for the loop to survive depercolation, combined with our observations that the loop does survive depercolation, indicate that the edge valence of 4.7 reported by He et al \cite{he2023single}. may be lower than the \textit{in vivo} redundancy due to sample preparation effects.

\section{Acknowledgements} This work was supported by the National Science Foundation (Grant No. 2105113). 

\bibliographystyle{unsrt}
\bibliography{edgerefs}

\end{document}